\documentclass[conference]{IEEEtran}
\IEEEoverridecommandlockouts

\usepackage{cite}
\usepackage{amsmath,amssymb,amsfonts}
\usepackage{algorithmic}
\usepackage{graphicx}
\usepackage{textcomp}
\usepackage{color,xcolor}
\def\BibTeX{{\rm B\kern-.05em{\sc i\kern-.025em b}\kern-.08em
    T\kern-.1667em\lower.7ex\hbox{E}\kern-.125emX}}

\newcommand{\BOS}{\textless BOS\textgreater}
\newcommand{\EOS}{\textless EOS\textgreater}
\newcommand{\SEP}{\textless SEP\textgreater}
\newcommand{\PAD}{\textless PAD\textgreater}
\newcommand{\UNK}{\textless UNK\textgreater}
\usepackage[misc,geometry]{ifsym}
\usepackage{url}
\usepackage{multirow}
\usepackage{setspace}
\usepackage{algorithm}
\usepackage[caption=false]{subfig}
\usepackage{booktabs}
\usepackage[hidelinks]{hyperref}
\usepackage{stfloats}
\usepackage{soul}
\usepackage{ulem}
\newcommand\blfootnote[1]{%
  \begingroup
  \renewcommand\thefootnote{}\footnote{#1}%
  \addtocounter{footnote}{-1}%
  \endgroup
}

\begin{document}

\title{PagPassGPT: Pattern Guided Password Guessing via Generative Pretrained Transformer}

\author{
\IEEEauthorblockN{Xingyu Su\textsuperscript{1,2}, Xiaojie Zhu\textsuperscript{3(\Letter)}, Yang Li\textsuperscript{1,2}, Yong Li\textsuperscript{2}, Chi Chen\textsuperscript{1,2}, Paulo Esteves-Veríssimo\textsuperscript{3}}
\IEEEauthorblockA{
\textit{School of Cyber Security, University of Chinese Academy of Sciences, Beijing, China} \\
\textit{Institute of Information Engineering, Chinese Academy of Sciences, Beijing, China} \\
\textit{King Abdullah University of Science and Technology, Thuwal, Kingdom of Saudi Arabia  } \\
\{suxingyu, liyang8119, liyong, chenchi\}@iie.ac.cn \\
\{xiaojie.zhu, paulo.verissimo\}@kaust.edu.sa
}
\thanks{(\Letter): Corresponding author}
}


\maketitle

\newcommand{\x}{\color{red}}
\newcommand{\xx}{\color{blue}}
\thispagestyle{plain}

\begin{abstract}

Amidst the surge in deep learning-based password guessing models, challenges of generating high-quality passwords and reducing duplicate passwords persist. 
To address these challenges, we present PagPassGPT, a password guessing model constructed on a Generative Pretrained Transformer (GPT).
 It can perform pattern guided guessing by incorporating 
 pattern structure information as background knowledge, resulting in a significant increase in the hit rate.
 Furthermore, we propose D\&C-GEN to reduce the repeat rate of generated passwords, which adopts the concept of a divide-and-conquer approach.
 The primary task of guessing passwords is recursively divided 
 into non-overlapping subtasks.
 Each subtask inherits the knowledge from the parent task and predicts succeeding tokens. 
In comparison to the state-of-the-art model, our proposed scheme exhibits the capability to correctly guess 12\% more passwords while producing 25\% fewer duplicates.

\end{abstract}

\begin{IEEEkeywords}
password guessing, generative pretrained transformer, trawling attack
\end{IEEEkeywords}

\section{Introduction}
\label{sec:introduction}

As we embrace the digital age, passwords have become ubiquitous in our society.
Accompanying the widespread use of passwords, the risk of password cracking is becoming a public concern. 
This threat arises from users' tendency to select meaningful characters as passwords~\cite{narayanan2005fast},
inadvertently making them susceptible to password guessing attacks, particularly targeted attacks and trawling attacks. 
Targeted attacks aim to crack users' passwords by collecting personally identifiable information and user identification credentials while trawling attacks focus on discovering user accounts that match known passwords~\cite{yu2022gnpassgan}.
\blfootnote{Our code is available at \url{https://github.com/Suxyuuu/PagPassGPT}.}

Flor{\^e}ncio \textit{et al.}~\cite{florencio2014administrator} investigated various user accounts and observed that the majority of accounts are not essential. 
According to their research, users might opt to create a new account rather than spend 10 minutes recovering a lost one. 
Rather than being a target, users are more likely to face threats from trawling attacks.

To enrich the literature on trawling attacks, extensive research has been conducted. 
In 1979, Morris \textit{et al.}~\cite{1979rule} proposed heuristic rules for generating passwords using dictionary words and utilized them in password guessing attacks.
Subsequently, traditional probabilistic models emerged and evolved, such as Probabilistic Context-Free Grammar (PCFG) models~\cite{PCFGModel-1, PCFGModel-2,PCFGModel-3,PCFGModel-4}, and Markov models~\cite{markov-1,markov-omen}. These models heavily depend on the training set and pose challenges in terms of generalization. To mitigate this concern, deep learning models are gradually introduced into the field of password guessing. 
Models, such as those based on Long Short-Term Memory (LSTM)~\cite{lstm,fla}, Generative Adversarial Network (GAN)~\cite{gan,wassersteingans,ganmodel,passgan}, Autoencoder (AE)~\cite{gan_vae_model,vaemodel-1,vaemodel-2},
and Generative Pretrained Transformer (GPT)~\cite{gpt,GPT2,gpt3,passgpt}, have contributed significantly to the advancement of password guessing models.
Particularly, PassGPT ~\cite{passgpt}, introduced by Rando \textit{et al.} in 2023, stands out as the state-of-the-art model in deep learning-based password guessing. Before the introduction of our scheme, it had the highest hit rate in trawling attacks, leveraging the capabilities of GPT. 
The experimental results illustrate a significant improvement in our scheme compared to theirs, with an increase of 12\% in hit rate.

\subsection{Problems}
Despite advancements in using deep learning technology for password guessing,
two challenges remain open. The first challenge is to improve the quality of passwords generated in pattern guided guessing. 
The second challenge is to minimize the likelihood of generating duplicate passwords during the guessing process, i.e., reducing the repeat rate of the passwords generated.

\begin{figure}[tbp]
    \centering
    \includegraphics[width=0.9\columnwidth]{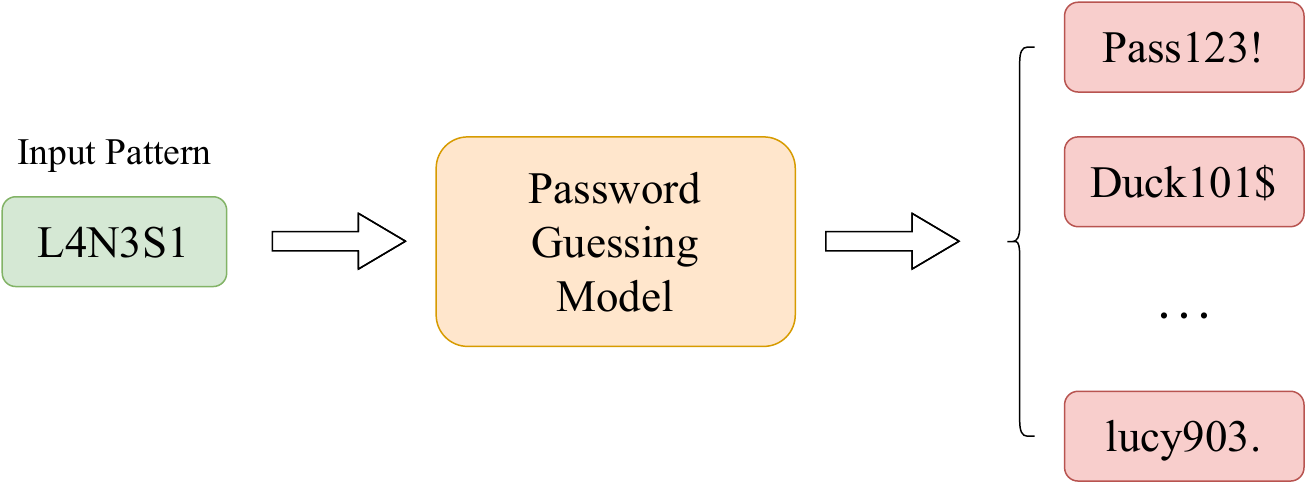}
    \caption{The process of pattern guided guessing. The pattern ``L4N3S1" signifies a password comprising four letters, followed by three numbers, and ending with one special character. Pattern guided guessing refers to the process wherein a password guessing model generates passwords that adhere to such specific patterns.}
    \label{fig:pattern guided guessing}
\end{figure}

\subsubsection{Pattern guided guessing}
A password guessing model is more effective when it possesses the ability to generate passwords guided by patterns. 
The process of pattern guided guessing is shown in Fig.\ref{fig:pattern guided guessing}.  
The success of traditional probabilistic models (e.g., PCFG models~\cite{PCFGModel-1,PCFGModel-2,PCFGModel-3,PCFGModel-4}) has validated that incorporating password patterns enhances a model's ability to crack more passwords. 
Furthermore, we have analyzed the pattern distribution in dozens of password datasets and observed a convergence in users' choice of password patterns. 
The top 10 patterns are consistent across all datasets and align with those observed within individual datasets. 
All these findings indicate that leveraging patterns from the known passwords as prior knowledge is a promising approach.

However, all the existing deep learning-based approaches do not support pattern guided guessing except PassGPT~\cite{passgpt}. 
PassGPT achieves pattern guided guessing by filtering candidate tokens during the token selection process based on a specific pattern. 
 Nevertheless, there is a high chance that the selected token deviates from the model's selection due to the specified pattern. 
For instance, the applied model predicts that the next token is a character while the pattern specifies a number, and finally, the scheme simply outputs a number without considering the model prediction. 
Due to the lack of consideration for the model's prediction,  this approach increases the likelihood of word truncation which contradicts the observations in~\cite{obser2006,obser2006-1,obser2010,obser2012,obser2014} that users are more likely to use meaningful words. 
This insight motivates us to integrate both model predictions and password patterns into the design of a new scheme.  In particular, we incorporate the password pattern as an initial condition in the password generation process, as illustrated in (\ref{eq:cs conditional probability}).

\subsubsection{Repeat rate}
Reducing the number of duplicate passwords can enhance the performance significantly.
Existing password guessing models generate each guess independently, akin to random sampling from the password space, leading to a large number of duplicate passwords.  
This issue is particularly pronounced with a large volume of guesses, resulting in a substantial number of duplicate passwords.
In our experiments, we find that PassGPT~\cite{passgpt}, the current state-of-the-art model, generates 34\% of duplicate passwords when making $10^9$ guesses. 
In slightly older models like PassGAN~\cite{passgan}, the repeat rate can be as high as 66\%, implying that over half of the guesses are useless.

\subsection{Our solutions}
\label{sec:our solutions}

\begin{figure}[tbp]
    \centering
    \includegraphics[width=0.95\columnwidth]{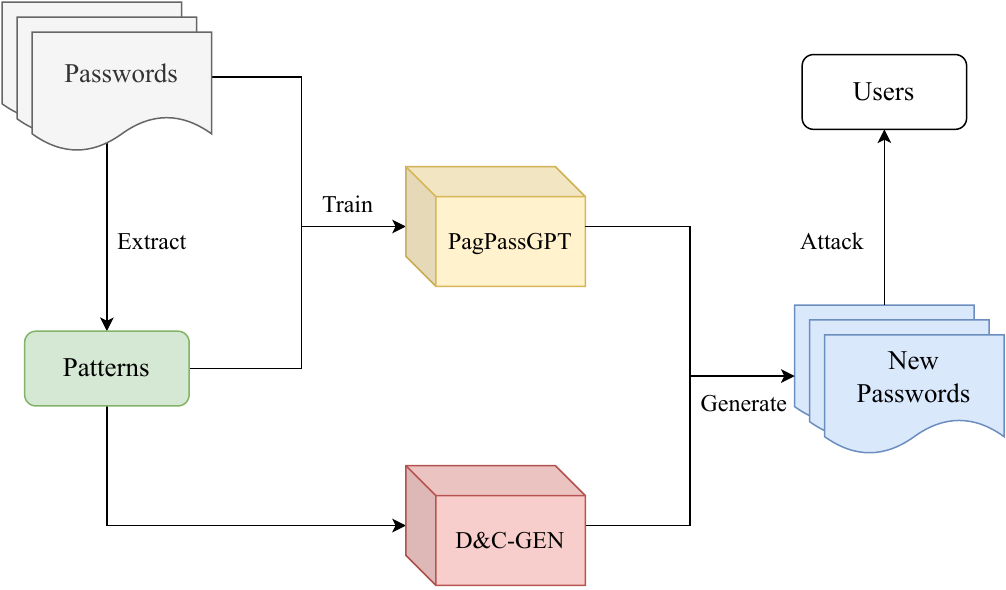}
    \caption{ The overview of the proposed solution. We utilize passwords and patterns extracted from known passwords to train PagPassGPT. Leveraging the ability of pattern guided guessing from PagPassGPT and with the assistance of D\&C-GEN, PagPassGPT generates high-quality passwords for trawling attacks.}
    \label{fig:overall}
\end{figure}

To address the aforementioned challenges, we propose a solution depicted in Fig.\ref{fig:overall}, comprising two core components: PagPassGPT, a password guessing model, and D\&C-GEN, a password generation algorithm. The former improves pattern guided password guessing, and the latter reduces the repeat rate during password generation. With the assistance of D\&C-GEN, PagPassGPT not only achieves higher hit rates but also maintains lower repeat rates.

\subsubsection{PagPassGPT}
The integration of model predictions and password patterns can be achieved by treating the preset password pattern as background knowledge during model predictions. It takes into account not only the constraint on the password pattern but also the model prediction. 
In a more formal representation, it is denoted by $\Pr(t_1, \cdots, t_n | P)$, where $t_i$ ($1 \leq i \leq n$) represents a token comprising the password, and $P$ stands for a password pattern.
In an auto-regressive model, tokens are generated sequentially. To match the mechanism, as illustrated in \eqref{eq:cs conditional probability}, we transform the conditional probability into an auto-regressive form. 
Based on that, we propose a design that adopts an auto-regressive model based on the second generation of Generative Pre-trained Transformer (GPT-2)~\cite{GPT2}, named PagPassGPT. 

\begin{equation}
     \label{eq:cs conditional probability} 
      Pr(t_1,t_2,\ldots,t_n | P) = \prod_{i=1}^n Pr(t_i | P, t_1, t_2, \ldots, t_{i-1})
\end{equation}

Specifically, we encode the password pattern information as the previous tokens preceding the password tokens and the model calculates the probability of the next token at each step based on the known tokens. PagPassGPT successfully achieves our goal of effectively generating passwords in a pattern guided guessing manner while also leveraging the power of GPT-2. 
In our experiments, we compared our scheme with PassGPT, and the results demonstrate that our scheme achieves up to approximately  27.5\% improvement in hit rate during the test of pattern guided guessing. 

The methodological distinction between PassGPT and the proposed PagPassGPT lies in their approach to utilizing password patterns. PassGPT strictly adheres to the password pattern by sequentially checking each element, whereas PagPassGPT derives the generated password based on the initial condition of the password pattern. 

As an example, consider the password pattern ``L1N1", denoting a letter followed by a number. In the PassGPT approach, the initial step entails selecting the character with the highest probability, followed by choosing a number with the highest probability under the condition of the previously selected character. Conversely, PagPassGPT starts by selecting the first character based on the ``L1N1" condition, ensuring it has the highest probability under this specific condition. Subsequently, the next number is chosen, taking into account both the ``L1N1" condition and the character chosen in the preceding step.

\subsubsection{D\&C-GEN}
After analyzing deep learning-based password guessing schemes, we observed that these schemes lack background knowledge during password generation as each of them applies almost the same initial environment, leading to a large number of duplicate passwords. 
To reduce the repeat rate of PagPassGPT,  inspired by the concept of the divide-and-conquer approach~\cite{Divide-and-conquerwiki},
we propose D\&C-GEN that recursively divides the main guessing task into small, non-overlapping subtasks with distinct requirements, including different patterns and different prefixes. For instance, one subtask may necessitate passwords conforming to the pattern ``L4N1" with the prefix ``abc", while another subtask may require passwords conforming to ``L1N4" with the prefix ``A12". 
Each subtask inherits all the requirements
from the parent task as the background knowledge and is intentionally designed to have no overlaps, resulting in a low repeat rate.
In the experiment, when the number of guesses reaches $10^9$,  the repeat rate of our proposed scheme is only 9.28\%, while 
PassGPT reaches 34.5\%.

Overall, the contributions of this paper are summarized as follows.

\begin{itemize}
    \item  We investigate the issue of trawling attacks and unveil the shortcomings of existing deep learning-based password guessing schemes. Additionally, we introduce PagPassGPT, which addresses these weaknesses by properly integrating deep learning-based models with password patterns. Furthermore, in an effort to reduce the occurrence of duplicate passwords, we propose the D\&C-GEN algorithm, which adopts a divide-and-conquer approach for the task of password guessing.
    \item We conduct thorough experiments to assess the effectiveness of PagPassGPT and D\&C-GEN on public datasets, performing a comparative analysis with the state-of-the-art models. Furthermore, we analyze the experimental results and conclude that the proposed schemes exhibit superior performance. 
\end{itemize}

\section{Background and Related Work}
\label{sec:Background and related work}

\subsection{Password Guessing}
Password guessing can be briefly categorized into two types based on whether the attack target is known: trawling attacks~\cite{1979rule,trawling} and targeted attacks~\cite{targeted-1,targeted-2,targeted-3}. These two attacks have different usage scenarios and evaluation strategies. 

\subsubsection{Trawling Attack} Trawling Attack~\cite{1979rule} is one of the earliest attacks that has drawn substantial attention. 
In a trawling attack, the attacker does not specifically target an individual user but rather focuses on a broad group of users.
The attacker builds password guessing models by modeling real leaked passwords~\cite{enwiki:dataleak,techrepublic2021}, and then uses the models to generate a large number of passwords to hit the new users' real passwords.
The attacker does not care about which user is under attack. 
The attacker's objective is to maximize the hit rate while minimizing the number of guesses.
Hence, password guessing models should prioritize generating passwords with higher probabilities of being used.

\subsubsection{Targeted Attack} The objective of a targeted attack is to rapidly crack the password of a specific user. 
Thus, the attacker would use personally identifiable information (PII)~\cite{enwiki:PII} or previously used passwords to launch an attack. 
In 2015, a targeted guessing model~\cite{wang2015emperor} based on Markov~\cite{markovwiki} was proposed. 
The main observation of this model is that users prefer to choose passwords based on names. 
After that, various models~\cite{targeted-1,targeted-2,targeted-3} are built based on PII or used passwords.

\subsection{Password Guessing Models for Trawling Attacks}
\label{sec:password guessing model}
Password guessing models are the core of password guessing. 
Extensive research has been conducted in this domain, categorizing the models into three types based on their technical foundations, presented chronologically as follows: rule-based models, probability-based models, and deep learning-based models.

\subsubsection{Rule-based Models}
The earliest models were rule-based models exemplified by tools like Hashcat~\cite{hashcatweb} and John the Ripper~\cite{johntheripperweb}. 
Both of them can perform rule-based attacks that output new passwords by applying transform rules to the old set of passwords. This approach is very fast but its shortcoming is obvious: it has a strong background knowledge dependency.

\subsubsection{Probability-based Models}
The probability-based models were proposed after rule-based models, such as Markov~\cite{markovwiki} models and Probabilistic Context-Free Grammar (PCFG)~\cite{PCFG} models. 
Narayanan \textit{et al.}~\cite{narayanan2005fast} proposed a single-layer Markov password guessing model in 2005, which used $n$-gram~\cite{ngram,enwiki:ngram}. 
It assumes that neighboring $n$ characters have a strong correlation and uses $n-1$ preceding characters to predict the next character. 
After that, Markus \textit{et al.}~\cite{markov-omen} proposed OMEN to improve the performance of Markov models.
In 2009, Weir \textit{et al.}~\cite{PCFGModel-1} proposed the first PCFG model for automated password guessing. 
After that,  various techniques are proposed to enhance the performance of PCFG models, such as adding new rules~\cite{PCFGModel-2}, introducing semantic information~\cite{PCFGModel-3}, and supporting long passwords~\cite{PCFGModel-4}.
However, all of the probability-based models have a common weakness that password guessing relies on a fixed vocabulary, which limits the diversity of generated passwords.

\subsubsection{Deep Learning-based Models}
With the emergence of deep learning, many deep learning-based techniques are applied to password guessing models.
In 2017, Melicher \textit{et al.}~\cite{fla} proposed FLA based on Long Short-Term Memory (LSTM)~\cite{lstm}, which is one of the first to introduce deep learning into password guessing. 
After that,  Recurrent Neural Networks (RNN)~\cite{rnn}, Generative Adversarial Networks (GAN)~\cite{gan,wassersteingans},  and Autoencoders (AE)~\cite{enwiki:autoencoder} are widely applied in this field.
Hitaj \textit{et al.}~\cite{passgan} proposed PassGAN in 2019, 
and Pasquini \textit{et al.}~\cite{gan_vae_model} developed a new framework named Dynamic Password Guessing (DPG) by using Wasserstein Autoencoders (WAE)~\cite{tolstikhin2018wasserstein} in 2021. 
One year later,  Yang \textit{et al.}~\cite{vaemodel-1} proposed VAEPass based on Variational Autoencoder (VAE)~\cite{vae}.
All these works have demonstrated the potential application of GAN and AE in this field. However, a challenge persists, namely the accuracy loss resulting from the mapping from continuous space to discrete space.
The latest trend in this field involves the utilization of language models, such as PassGPT~\cite{passgpt} based on Generative Pretrained Transformer (GPT) and PassBERT~\cite{passbert} based on Bidirectional Encoder Representations from Transformers (BERT)~\cite{bert}.
These models provide an approach by treating passwords as short texts.

\subsection{Probabilistic Context-Free Grammar Models}
\label{sec:Probabilistic Context-Free Grammar}
Probabilistic Context-Free Grammar (PCFG)~\cite{PCFG} extends Context-Free Grammar (CFG)~\cite{cfg}, providing a framework for describing the syntactic structure of sentences in natural language with the incorporation of probabilities.
PCFG models~\cite{PCFGModel-1,PCFGModel-2,PCFGModel-3,PCFGModel-4} fall under the category of password guessing models that integrate PCFG with passwords.
In 2009, Weir \textit{et al.}~\cite{PCFGModel-1} proposed the first scheme based on PCFG. The core concept is to divide passwords into segments based on character types (letters, numbers, and special characters). 
Throughout the training process, the model computes and retains the probabilities associated with patterns and segments.
In the generation process, it prioritizes patterns based on their probabilities. For each pattern, it selects segments in descending order of probability that adhere to the pattern.
For instance, given the password ``abc123!", the scheme first divides it into three segments (``abc", ``123", and ``!"), and then the entire password pattern is represented as ``L3N3S1". ``L3", ``N3", and ``S1" represents three letters, three numbers, and one special character respectively. The probability of the entire password can be expressed as follows:
\begin{align}
\label{eq:PCFG}
    Pr(abc123!)=&Pr(L3N3S1) \cdot Pr(abc|L3) \notag \\
    &Pr(123|N3) \cdot Pr(!|S1)
\end{align}

Subsequent PCFG models have introduced various enhanced techniques, including improvements in password segmentation methods~\cite{PCFGModel-2, PCFGModel-3} and enhancements in the capability to generate longer passwords~\cite{PCFGModel-4}. 
Despite numerous improvements in subsequent research, two primary challenges persistently remain unresolved. The first challenge is that PCFG models struggle to generate words that are not present in the vocabulary. The second challenge involves the difficulty of perfectly segmenting passwords into appropriate segments.

\subsection{Generative Pretrained Transformer}
\label{sec:Generative Pretrained Transformer}
Generative Pretrained Transformer (GPT)~\cite{gpt,GPT2,gpt3} is a series of natural language processing models proposed by OpenAI~\cite{openaiweb}, which has excellent text generation capabilities.
It undergoes pre-training through unsupervised learning on a broad corpus of diverse textual data. Its core architecture comprises multiple layers of transformer decoders~\cite{transformer}. These layers leverage attention mechanisms~\cite{transformer} for efficient feature extraction and parallel processing of data.

GPT's generation is accomplished through a process called ``auto-regression"~\cite{autoregressive}. During text generation, GPT processes input tokens, incorporating information from preceding tokens to predict the likelihood of the next token. This process iterates sequentially, with each token generated based on the context established by the preceding tokens.
Therefore, the probability of generating a sequence of tokens can be represented as below:
\begin{equation}
    Pr(x_1,x_2,\ldots,x_n) = \prod_{i=1}^n Pr(x_i | x_1, x_2, \ldots, x_{i-1})
\end{equation}

In contrast to n-gram-based models, auto-regressive generation utilizes all preceding tokens for prediction. This mechanism not only enhances the model's comprehension but also allows for better control of the generation process by manipulating input tokens.

\begin{figure*}[htb]
    \centering
    \includegraphics[width=1.95\columnwidth]{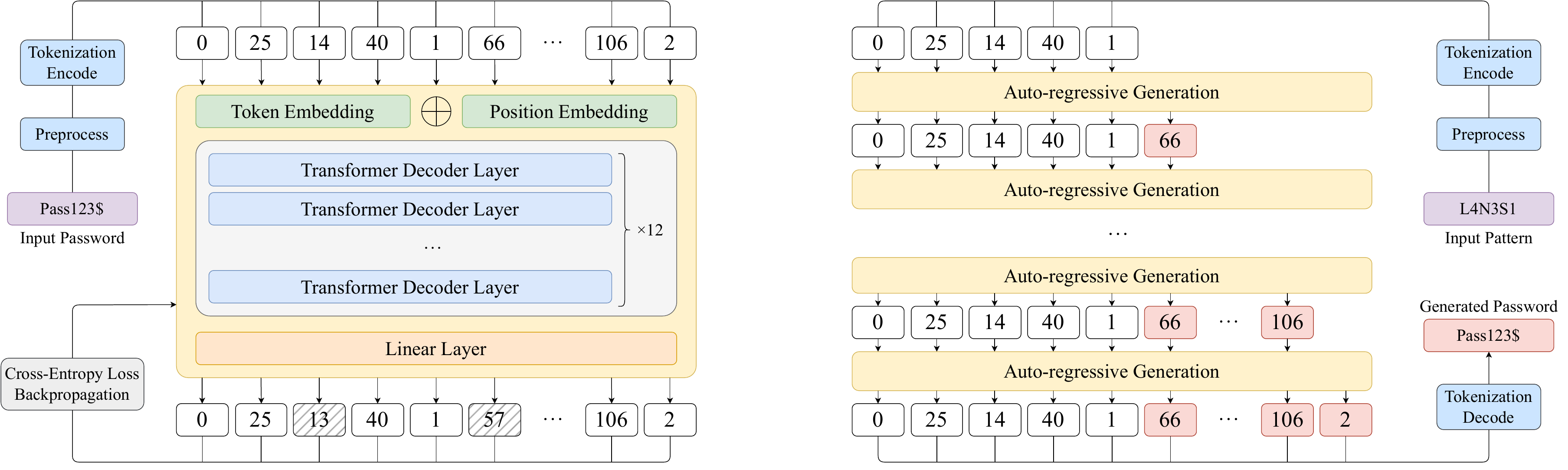}
    \caption{The training process (left) and the generation process (right) of \textit{PagPassGPT}. 
    The numbers in the figure correspond to the indexes after encoding, as presented in Fig. {\ref{fig:encode and decode}}. Instances, where the number is shadowed, denote incorrect predictions, while numbers highlighted in red signify predicted indexes of a new password.}
    \label{fig:model}
\end{figure*}

\section{Our Approach}
\label{sec:Our Approach}

In this section, we start by introducing the threat model. Subsequently, we present our proposed scheme, PagPassGPT. Finally, we illustrate its enhancement algorithm, D\&C-GEN, designed to reduce duplicate passwords.

\subsection{Threat Model}
In this paper, following{~\cite{passgpt}}{~\cite{xu2021chunk}},  we concentrate on trawling attacks as the targeted threat model. In trawling attacks, the assailant endeavors to recover passwords by making an extensive number of guesses, such as up to $10^{14}${~\cite{florencio2016pushing}}{~\cite{tan2020practical}} as the upper limit. This choice of a large number of guesses aligns with a practical attacker scenario considering the available computing power.

\subsection{PagPassGPT}
\label{sec:PagPassGPT}

As shown in  Fig. \ref{fig:model},  PagPassGPT includes two parts, training and generation. 
In the phase of training, the input is the passwords that are from the training set and the output is the trained model.
During the generation phase, the input is the password pattern that is applied to guide the password generation, 
and the output is the generated passwords. 

Following PassGPT{~\cite{passgpt}}, PagPassGPT is built upon GPT-2{~\cite{GPT2}}, which is the second generation of the GPT model introduced by OpenAI. 
GPT-2 is known for its open-source nature and robust generative capabilities. 
As discussed in Section {\ref{sec:our solutions}}, we need an auto-regressive model to calculate the conditional probability of passwords in pattern guided guessing.
GPT-2, being a decoder-only model employing masked self-attention mechanisms, excels in generative tasks by considering all preceding tokens when generating new ones.
Moreover, in comparison to other models such as GAN{~\cite{gan}}, VAE{~\cite{vae}}, and flow-based models{~\cite{enwiki:flow}}, GPT-2 is particularly well-suited for learning the intrinsic characteristics of discrete texts{~\cite{surveyGAN}}. Even when compared with LSTM{~\cite{lstm}}, another text model based on deep learning, GPT-2 exhibits a superior parallel mechanism, allowing for faster training{~\cite{surveyTransformers}}. Additionally, thanks to its attention mechanism, GPT-2 demonstrates stronger semantic understanding and more robust feature extraction capabilities{~\cite{researchLSTM}}.

\subsubsection{Training Process}
\begin{figure}[tbp]
    \centering
    \includegraphics[width=0.95\columnwidth]{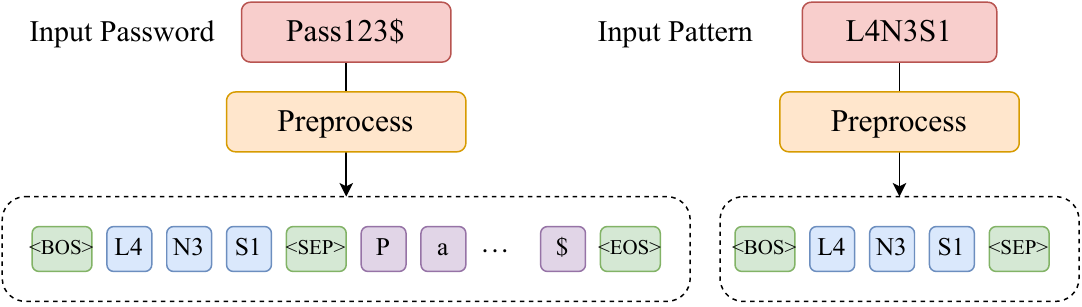}
    \caption{The preprocessing operation of tokenizer of \textit{PagPassGPT}. 
    On the left side, it shows that during the training phase, the password pattern is preprocessed and outputs the concatenation of the password pattern and password with a
    format, named rule. 
    On the right side, it shows that during the generation phase, the input of the password pattern is preprocessed into another short rule that is ready to be embedded. 
    }
    \label{fig:preprocess}
\end{figure}

\begin{figure}[tbp]
    \centering
    \includegraphics[width=0.9\columnwidth]{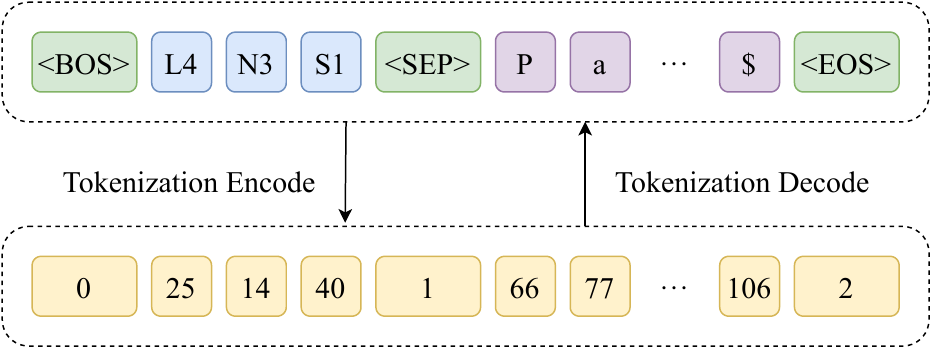}
    \caption{The tokenization process of the tokenizer of \textit{PagPassGPT} contains two functions: encode and decode. The encode takes a rule as input and produces tokenized indexes while decoding reverses the process.}
    \label{fig:encode and decode}
\end{figure}

As shown in the left part of Fig. \ref{fig:model},
during the training process, each input password undergoes preprocessing followed by tokenization, 
both implemented within a specialized component called the tokenizer.
As shown in the left part of Fig. \ref{fig:preprocess}, 
in the phase of training preprocessing, 
the tokenizer of PagPassGPT applies PCFG (detailed in Section \ref{sec:Probabilistic Context-Free Grammar}) to extract the password pattern, ``L4N3S1",  from the input password, ``Pass123\$".
 ``L4N3S1" stands for a password consisting of four letters followed by three numbers and one special character.
Then, it utilizes the extracted pattern to concatenate with the password, forming a rule in the format below, 
\begin{equation*}
    <BOS>||\ {Pattern}\ || <SEP> ||\ {Password}\ || <EOS>
\end{equation*}
where \BOS~represents the beginning of the sequence, \SEP~denotes the separator, and \EOS~stands for the ending of the sequence.
In the phase of tokenization, as shown in Fig. \ref{fig:encode and decode},  the tokenizer serves two functions: encoding and decoding. 
During encoding, the tokenizer takes the former rule as input and produces tokenized indexes, while during decoding, it reverses the process, mapping tokenized indexes back to the rule. 
In particular, during the encoding phase, the tokenizer initially splits the input content into segments, considering each segment as a token. 
Each token is then mapped to an index based on a vocabulary, ensuring that every token has a unique index.
For example, as shown in Fig. \ref{fig:encode and decode}, 
the tokenizer splits the input rule into segments, and each of them is mapped to an index according to the applied vocabulary, forming the index list $[0,  25,  14,  40,   1,  66,  77,  95,  95,  42,  43,  44, 106,   2]$.
The applied vocabulary consists of three categories of tokens: 94 visible ASCII character tokens, excluding the space character; 5 special tokens (\BOS, \SEP, \EOS, \UNK, and \PAD); and 36 pattern tokens (e.g., L12, S12, and N12), totaling 136 tokens.
\UNK~is a token used to represent an out-of-vocabulary token, and \PAD~is the padding token utilized to pad the indexes list.

After tokenization of the training phase, the tokenized indexes
are taken as the input of the embedding process consisting of token embedding and position embedding~\cite{bert}. The two processes are implemented by two linear layers and their outputs will be added together.

After that, the embedded result is input to the 12 Transformer decoder layers~\cite{transformer} similar to GPT-2 architecture~\cite{GPT2}.
Finally, through a linear layer named language modeling head, 
it outputs a probability distribution over the vocabulary using the Softmax function 
which is optimized by reducing the cross-entropy iteratively during the whole training process.

 \subsubsection{Generation Process}
For the generation process, the input is a pattern. 
As shown in the right part of Fig. \ref{fig:preprocess}, the input pattern is first transformed into a format as below and then tokenized into indexes by the tokenizer.
\begin{equation*}
\label{eq:generation}
    <BOS> ||\ {Pattern}\ || <SEP>
\end{equation*}
Particularly, with the encoded initial pattern as input, it 
predicts the index recursively based on both the pattern information and the history of the generated index. 
As shown in the right part of Fig. \ref{fig:model}, the initial index list is $[0, 25, 14, 40, 1]$ and the first predicted index is $66$ based on the index list. 
The auto-regressive generation mechanism is invoked recursively until all the indexes are generated. After that, the newly generated indexes are decoded and output the guessed password.  
This approach enables the generation of high-quality passwords, which aligns with the semantic characteristics of passwords and password pattern requirements.

\subsection{D\&C-GEN}
\label{sec:Guessing Algorithm}

D\&C-GEN is proposed to reduce the repeat rate of generated passwords, inspired by the concept of the divide and conquer approach.  
In this section, we detail the D\&C-GEN algorithm in three parts. 
The first part is to illustrate its implementation and the second part is to analyze its effectiveness. 
Finally, we demonstrate its optimization.

\subsubsection{Design}
\begin{figure}[tbp]
    \centering
    \includegraphics[width=0.6\columnwidth]{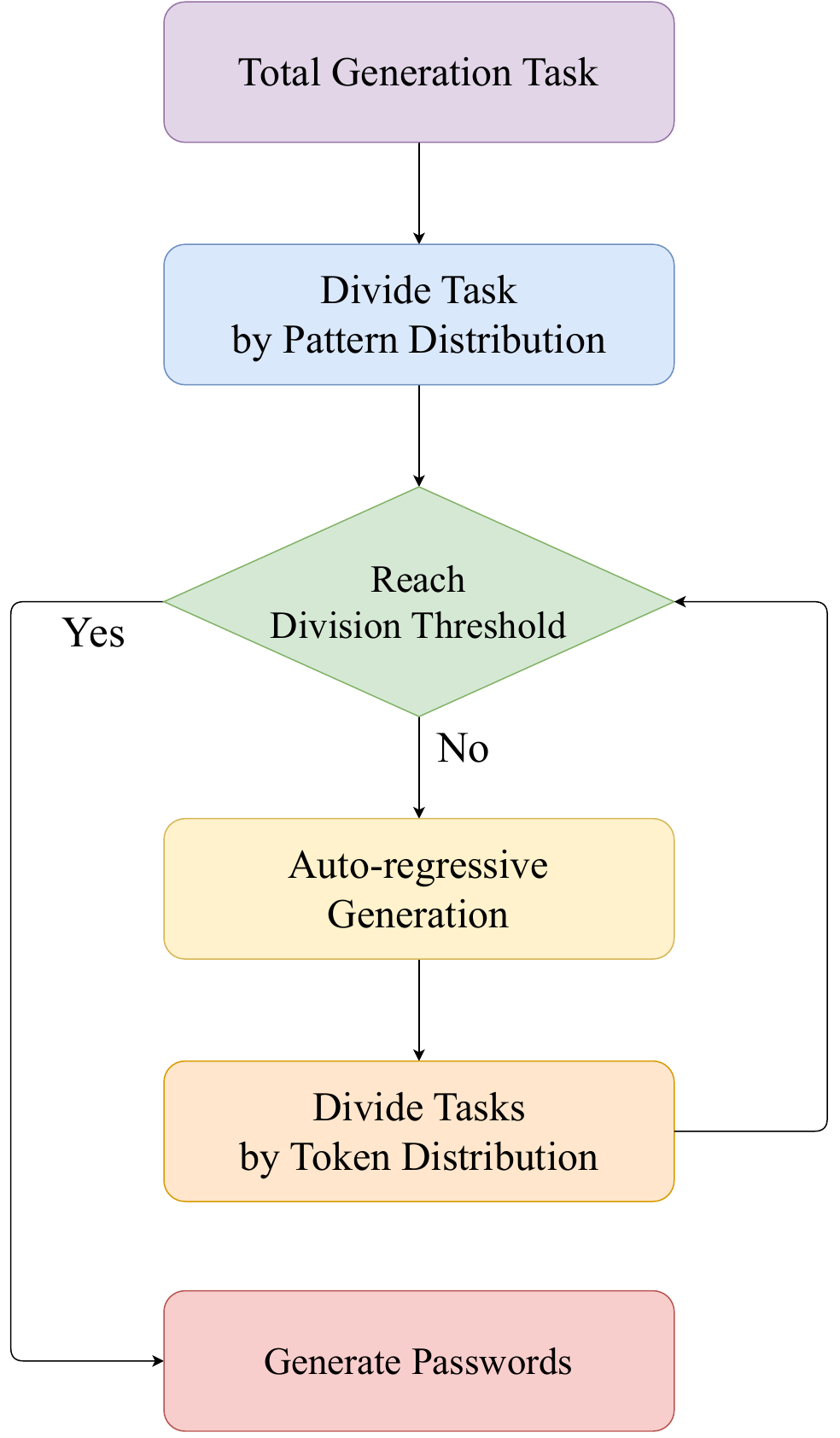}
    \caption{The workflow of D\&C-GEN's process.  The whole generation task is first divided into numerous subtasks based on the pattern distribution. If a subtask reaches the division threshold, it is executed to generate passwords. Otherwise, it undergoes auto-regressive generation and is further divided by the new token distribution into more subtasks.
    }
    \label{fig:algorithm structure}
\end{figure}

As shown in Fig. \ref{fig:algorithm structure}, the workflow of the D\&C-GEN starts from a task, and then this task is recursively split into many subtasks. 
In addition, a threshold is set to control the granularity of the division. 
If the threshold is reached, the division job is stopped and followed by password generation. 
In detail, a pattern is first selected from the pattern space and then its corresponding probability 
is read. Based on the pattern probability, the number of passwords to be generated is computed by using the total number of guessed passwords multiplying the probability. If the result is smaller than the threshold, then the task is executed to generate passwords under its requirements.
Otherwise, the task is added into a list to prepare for further division. 
For each element of the list, it first evaluates whether the number of passwords to be generated is larger than the threshold. 
If it is larger than the threshold, it is executed to generate the following token based on the current prefix,  resulting in subtasks with longer prefixes. 
Those newly generated subtasks are added back to the list.

\begin{table}[tbp]
    \centering
    \caption{The frequently used notations.}
    \label{tab:notation}
    \begin{tabular}{cl}
     \toprule
       \textbf{Notation}  & \textbf{Description}  \\
       \midrule 
       $R$ & The set of generated passwords \\
       $T$  &  The threshold of dividing a task \\ 
       $N$ & The total number of  guesses  \\
       $S_p$ & The set of patterns and their probabilities \\ 
       \multirow{2}{*}{\textit{Tokens}} & The set of candidate tokens and \\
       & their probabilities\\ 
       \textit{Pref} & The prefix used to generate passwords  \\
       \multirow{2}{*}{$N_{P_i}$} & The number of passwords to be generated\\ 
       & conforming to a pattern $P_i$ \\
       $n$ & The number of  passwords to be generated \\ 
       \bottomrule 
    \end{tabular}
\end{table}

\begin{figure}[tbp]
    \centering
    \includegraphics[width=0.9\columnwidth]{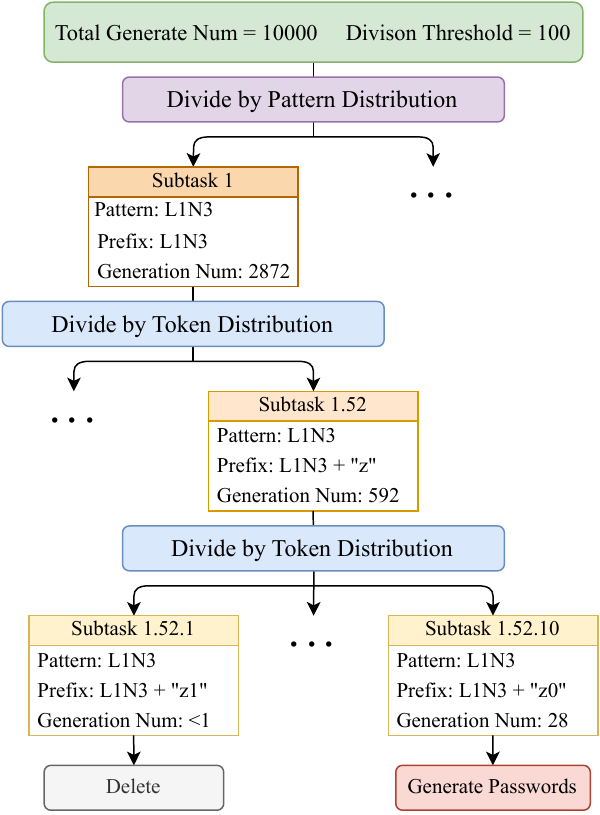}
    \caption{An instance of executing D\&C-GEN. 
    In this example, the total guessing number $N$ is set to 10,000, and the threshold $T$ of task division is set to $100$. 
    The initial task is divided into subtasks by pattern distribution, and each subtask has its prefix and its generation number (i.e., the number of passwords to be generated).
    If the generation number is less than $T$, this subtask is executed to generate passwords.
    If not, this subtask is further divided by the distribution of the next token.
    Especially, if the generation number is less than $1$, i.e., the probability of passwords with the current prefix is almost impossible, the subtask is deleted. It's worth noting that every division by token distribution is filtered according to the pattern requirement, e.g. the subtask 1.52 only has 10 new subtasks because the next token is expected to be a number.}
    \label{fig:algorithm example}
\end{figure}

\begin{algorithm}[tbp]
\setstretch{1.3}
	\renewcommand{\algorithmicrequire}{\textbf{Input:}}
	\renewcommand{\algorithmicensure}{\textbf{Output:}}
	\caption{D\&C-GEN}
	\label{Guessing Algorithm}
	\begin{algorithmic}[1]
		\REQUIRE $S_{p}=\left\{\ P_i, Pr(P_i)\ \vert \ i \in \left[1,m\right] \right\}$,  $T$, $N$

            \STATE $R = \phi$ 
            \FOR{$i = 1$ to $m$}
                \STATE  $N_{P_i} = N\cdot Pr(P_i)$
                \IF{$N_{P_i} \leq T$}
                    \STATE Generate $N_{P_i}$ passwords and add them into $R$
                \ELSE
                    \STATE  ${Pref}_0=<BOS>||\ P_i\ ||<SEP>$
                    \STATE  Initialize list $L_{P_i}$
                    \STATE Push $({Pref}_0,\ N_{P_i})$ to $L_{P_i}$
                    \WHILE{$L_{P_i}$ is not empty}
                    {
                        \STATE  Pop an element $({Pref}_i,\ n_i)$ from $L_{P_i}$
                        \IF{$n_i \leq T$}
                            \STATE Generate $n_i$ passwords and add them into $R$
                        \ELSE
                            \STATE Get $Tokens=\left\{\ t_j,Pr(t_j)\ \vert \ j \in \left[1,c\right] \right\}$ \\ calculated by model based on $Pref_{i}$
                            \FOR{$j=1$ to $c$}
                                \STATE  $n_j = Pr(t_j)\cdot n_i$
                                \STATE  ${Pref}_j = {Pref}_i\ ||\ t_j$
                                \STATE Push $({Pref}_j,\ n_j)$ to $L_{P_i}$
                            \ENDFOR
                        \ENDIF
                    }
                    \ENDWHILE
                \ENDIF
            \ENDFOR
         
		\ENSURE  Generated passwords set $R$
	\end{algorithmic}  
\end{algorithm}

The above description is detailed in Algorithm \ref{Guessing Algorithm}. 
It takes the total number $N$ of password guessing attempts, the threshold  $T$ of dividing a task, and a set $S_p$ of patterns and their probabilities.  
For each pattern $P_i$, we first compute the number $N_{P_i}$ of passwords to be generated through the total number $N$ of attempts multiplying the probability $Pr(P_i)$ of the pattern. 
After that, the comparison between the number of passwords to be generated and the threshold is conducted. 
If the number of passwords to be generated
is smaller than the threshold,  the task is directly executed and outputs the passwords to a set $R$. 
Otherwise, the pattern $P_i$ and the number $N_{P_i}$ of passwords to be generated conforming to the pattern $P_i$
is added into a list $L_{P_i}$. 
For each element ($\textit{Pref}_i$, $n_i$) of $L_{P_i}$, 
We first pop it from the list and then compare its number $n_i$ of passwords to be generated with the threshold $T$. 
If it is smaller than the threshold, the task is executed and outputs the passwords to $R$.
Otherwise, the model is executed to get the probability of following $c$ tokens with the current prefix $\textit{Pref}$.
$c$ is the number of candidate tokens conforming to the current pattern requirement. 
In our setup, the variable $c$ is assigned different values: $52$ for a letter, $10$ for a number, and $32$ for a special character, depending on the type of the next token. 
After that, for each new token, the number $n_j$ of passwords to be generated
is calculated and the current prefix is concatenated with the new token to form the new prefix $\textit{Pref}_j$. Subsequently,  the number $n_j$ of passwords to be generated
and the new prefix $\textit{Pref}_j$ are added into the 
list $L_{P_i}$. 
Finally, the algorithm outputs the set $R$. 
For clarity, we've included an example in Fig. \ref{fig:algorithm example} to illustrate the process of D\&C-GEN.

\subsubsection{Analysis}
From the above description of D\&C-GEN, 
we can learn that repeated passwords are only possibly generated in a single small subtask 
since multiple passwords may be generated at a time with the same prefix. 
There are no same passwords existing in different tasks. 
If $T$ is small, the chance of generating duplicate passwords is very low.
However, if $T$ is too small, a large number of tasks will be overloaded.   
It is crucial to carefully choose the threshold $T$, considering both computational and parallelization capabilities.

\subsubsection{Optimization}
To balance both the quality of generated passwords and guessing speed, we can optimize D\&C-GEN in the following aspects.

\begin{itemize}
    \item To reach the best utility of GPUs,   
     $T$ can be set to the maximum number of passwords that can be generated in parallel by a single GPU.
    \item Before the task is executed, the evaluation of $N_{P_i}$ is first conducted. If it exceeds the maximum number of passwords based on the pattern, the value should be reset to the maximum number. 
    For example, if a pattern is ``N3", then the maximum number of guesses is  1000. However,  
    if the computed $N_{P_i}$ is 5000, then it should be reset to 1000. It reduces the number of useless guesses. 
    \item To enhance efficiency, the tasks in the list can be executed concurrently.
    \item To reduce the frequency of encoding and decoding,  all prefixes can be stored as tensors. 
\end{itemize}

\section{Evaluation}
\label{sec:evaluation}

In this section, we first introduce the applied datasets and models. 
After that, we illustrate the experimental comparison results with the related work.

\subsection{Datasets}
\label{sec:datasets}
In the experiment, we adopt five datasets:  Rockyou~\cite{Rockyouwiki}, 
LinkedIn~\cite{linkedinwiki},
phpBB~\cite{phpbbleak},
MySpace~\cite{spaceleak}, 
and Yahoo!~\cite{Yahoowiki}.  
In total, there are 75,349,874 entries.
The applied datasets are consistent with PassGPT~\cite{passgpt}, except for the exclusion of the Hotmail dataset due to its small size. We opt for the Yahoo! dataset as a replacement, following the recommendation of Melicher \textit{et al.}~\cite{fla}.
The details of the adopted datasets are illustrated in Table \ref{tab:datasets}. The first two datasets, LinkedIn and Rockyou, are utilized for both training and testing purposes, while the remaining datasets are employed for cross-site evaluation.

\begin{table}[tbp]
    \centering
    \caption{Key characteristics of applied datasets. }
    \label{tab:datasets}    
    \begin{tabular}{lrrr}
    \toprule
    \textbf{Name} & \textbf{Unique} & \textbf{Cleaned}& \textbf{Retention rate}  \\ \midrule
    RockYou  &14,344,391  &13,265,184  & 92.5\% \\
    LinkedIn & 60,525,521 & 49,776,665 & 82.2\%     \\
    phpBB    & 255,376    & 251,283    & 98.4\%  \\
    MySpace  & 37,126     & 36,369     & 98.0\%      \\
    Yahoo!   & 442,836    & 436,015    & 98.5\%   \\ \bottomrule
    \end{tabular}
\end{table}

\subsubsection{Data Cleaning}
Aligned with the recommendations in~\cite{vaemodel-1, passgpt, passbert}, we conducted data cleaning, excluding excessively long and short passwords, and retained those with lengths ranging between 4 and 12 characters.
This approach takes into account both related works and the frequency analysis of password datasets. Outlier passwords constitute only a very small proportion of the dataset, and their presence does not significantly impact the evaluation results.
In addition, we removed all duplicate passwords and the passwords containing
Non-ASCII characters and invisible ASCII characters, retaining only digits, letters, and special characters (excluding the space character).

\subsubsection{Data Utilization}
The Rockyou and LinkedIn dataset is divided into training, validation, and test sets in a 7:1:2 ratio, respectively. The training and validation sets are used for model training, while the test set is reserved exclusively for evaluation. 
For the test of pattern guided guessing in Section {\ref{sec:pattern guided guessing}} and the test of trawling attack guessing in Section {\ref{sec:Trawling Attack Guessing}}, we use Rockyou only. For the test of cross-site attack in Section {\ref{sec:cross-site attack}}, both Rockyou and LinkedIn are used.
The three remaining datasets are employed entirely for cross-site evaluation.

\subsubsection{Ethical Claim}
We ensure the ethical foundation of our work through the following aspects:
\begin{itemize}
    \item Public data. All datasets are public on the Internet and we do not share them with others.
    \item Necessary data. We minimize data usage, utilizing only what is essential and necessary for the research.
    \item No additional harm. All data will be utilized solely for research purposes and will not be employed in practical real-world applications.
\end{itemize}

\subsection{Models}
\label{sec:Models}

\subsubsection{Our Model}
PagPassGPT was trained using a batch size of 512 for 30 epochs, employing the AdamW optimizer with an initial learning rate of 5e-5 through the GPT2 library~\cite{gpt2library}. The training process, conducted on a Linux system with four GeForce RTX 3080 GPUs, took over 25 hours.

The parameters of our model are demonstrated below:
\begin{itemize}
    \item Max number of input tokens: 32
    \item Embedding size: 256
    \item Number of hidden layers: 12
    \item Number of attention heads for each attention layer: 8
\end{itemize}

\subsubsection{Models for Comparison}
\label{sec:Models for Comparison}
In selecting the comparison model, we choose the most recent and relevant work. 
Especially, PassGAN~\cite{passgan} based on GAN, VAEPass~\cite{vaemodel-1} based on VAE, PassFlow~\cite{passflow} based on flow~\cite{flow}, and PassGPT~\cite{passgpt} based on GPT are chosen.
All the models for comparison are trained using the training sets that do not contain any passwords from the test set and are evaluated on the same test set. All configurations of models are consistent with the description of the original papers.

\subsection{Pattern Guided Guessing Test}
\label{sec:pattern guided guessing}
Given that only PassGPT can perform pattern guided guessing, our comparison evaluates PagPassGPT against PassGPT in the pattern guided guessing test.

The experiment of pattern guided guessing test is designed in five steps. 
The initial step involves computing the probability distribution of extracted patterns from passwords in the test set and categorizing them based on different numbers of segments.
Each segment represents a format for organizing characters. 

For example, the pattern of ``password123" is ``L8N3" and this pattern is classified into the category with two segments (i.e.,  ``L8" and ``N3").
The second step is to select the target patterns. 
In our experiment, we choose the twenty-one most frequent patterns within each category\footnote{ The reason why we decided twenty-one is that the category with the least patterns has only twenty-one patterns.}.
There are a total of twelve categories, ranging from one segment to twelve segments since the maximum length of a password is twelve after data cleaning.  
The third step is to run the model and output the generated passwords. 
In our setting, following the configuration of trawling attacks~\cite{passgpt,passflow,yu2022gnpassgan}, 
we execute PassGPT and PagPassGPT to generate 100,000 passwords for each target pattern. 
The last step is to calculate the hit rate. 
In particular, we evaluate both the hit rate $HR_s$ of a category with $s$ segments and the hit rate $HR_P$ of a specific pattern $P$ as below. 
\begin{align}
    {HR}_{s} = {NH}_s/{TC}^{test}_s \\
    {HR}_{P} = {NH}_P/{TC}^{test}_P
\end{align}
where $NH$ denotes the number of hits, ${TC}_s^{test}$ and ${TC}_P^{test}$ represent the total number of passwords from the test set conforming to a certain category with $s$ segments and conforming to a specific pattern $P$ respectively.

\begin{figure}[tbp]
    \centering
    \includegraphics[width=0.95\columnwidth]{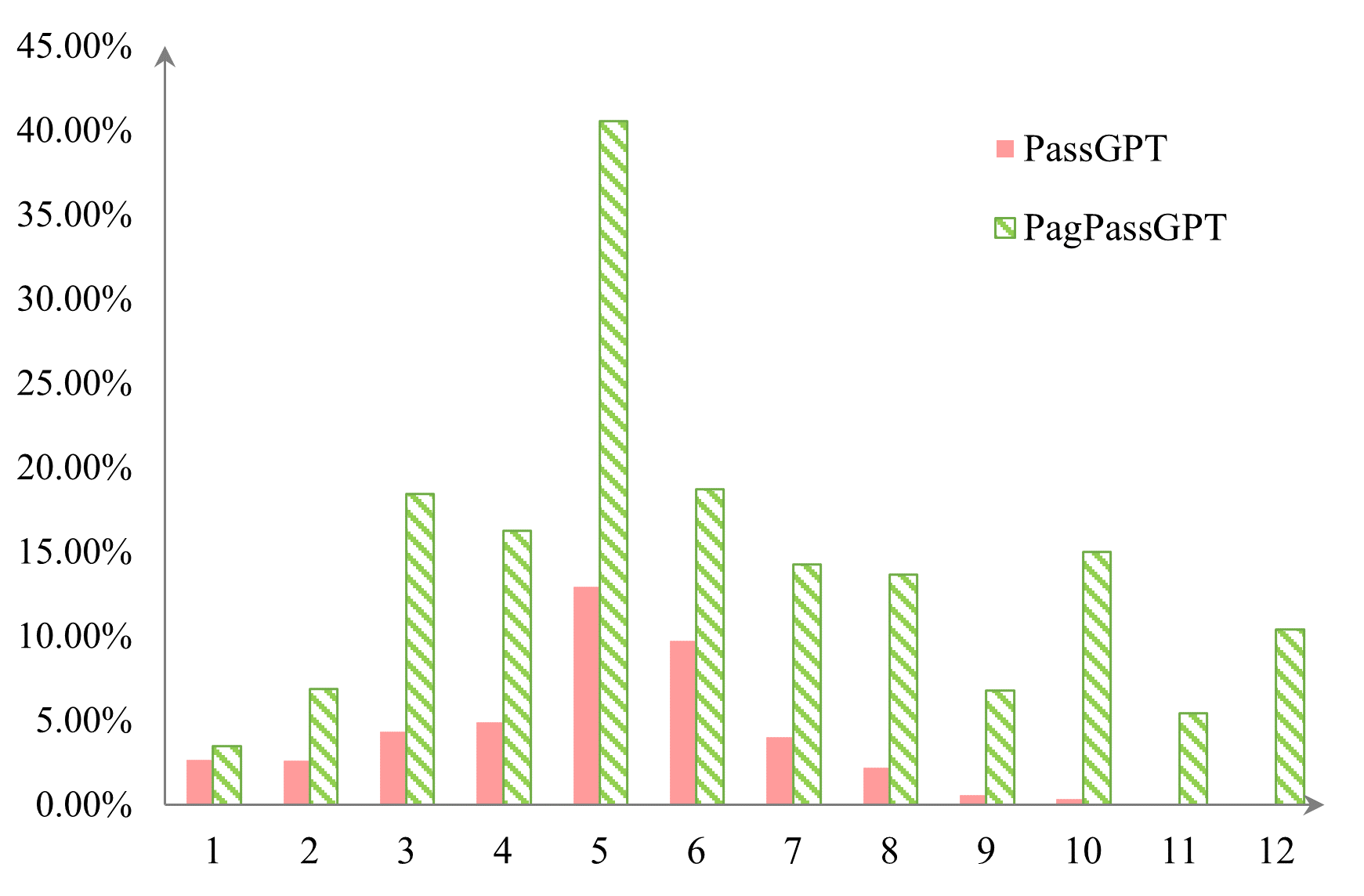}
    \caption{Compare $HR_s$ of PassGPT and PagPassGPT, $s\in [1,12]$. The vertical axis represents the hit rate $HR_s$, while the horizontal axis represents 
    categories with different numbers of segments.}
    \label{fig:chunksize=1-12}
\end{figure}

\begin{figure*}[tbp]
\centering
        \subfloat[]{\includegraphics[width=0.9\columnwidth]{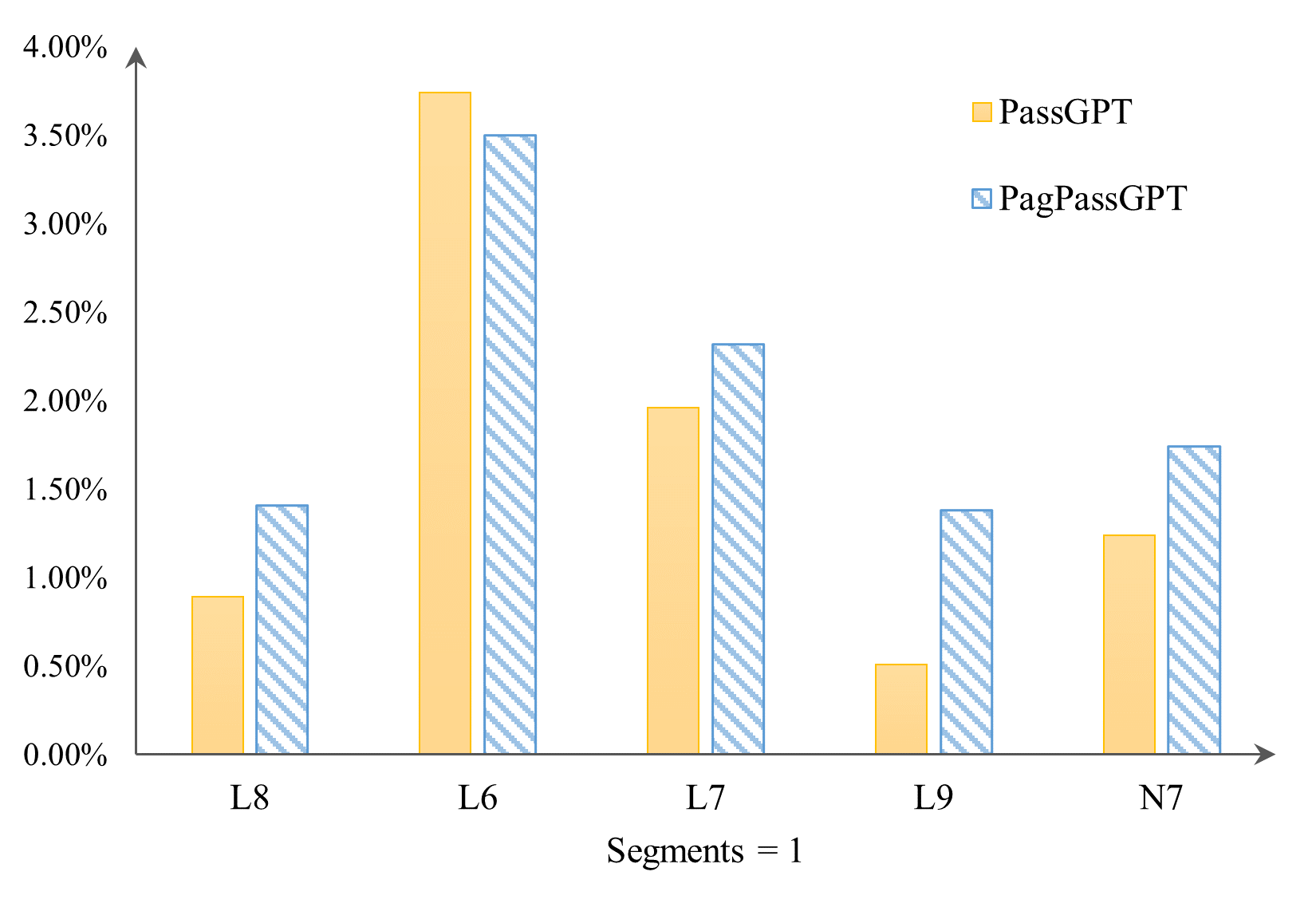}
        \label{fig:chunksize_1}}
        \hfil
        \subfloat[]{\includegraphics[width=0.9\columnwidth]{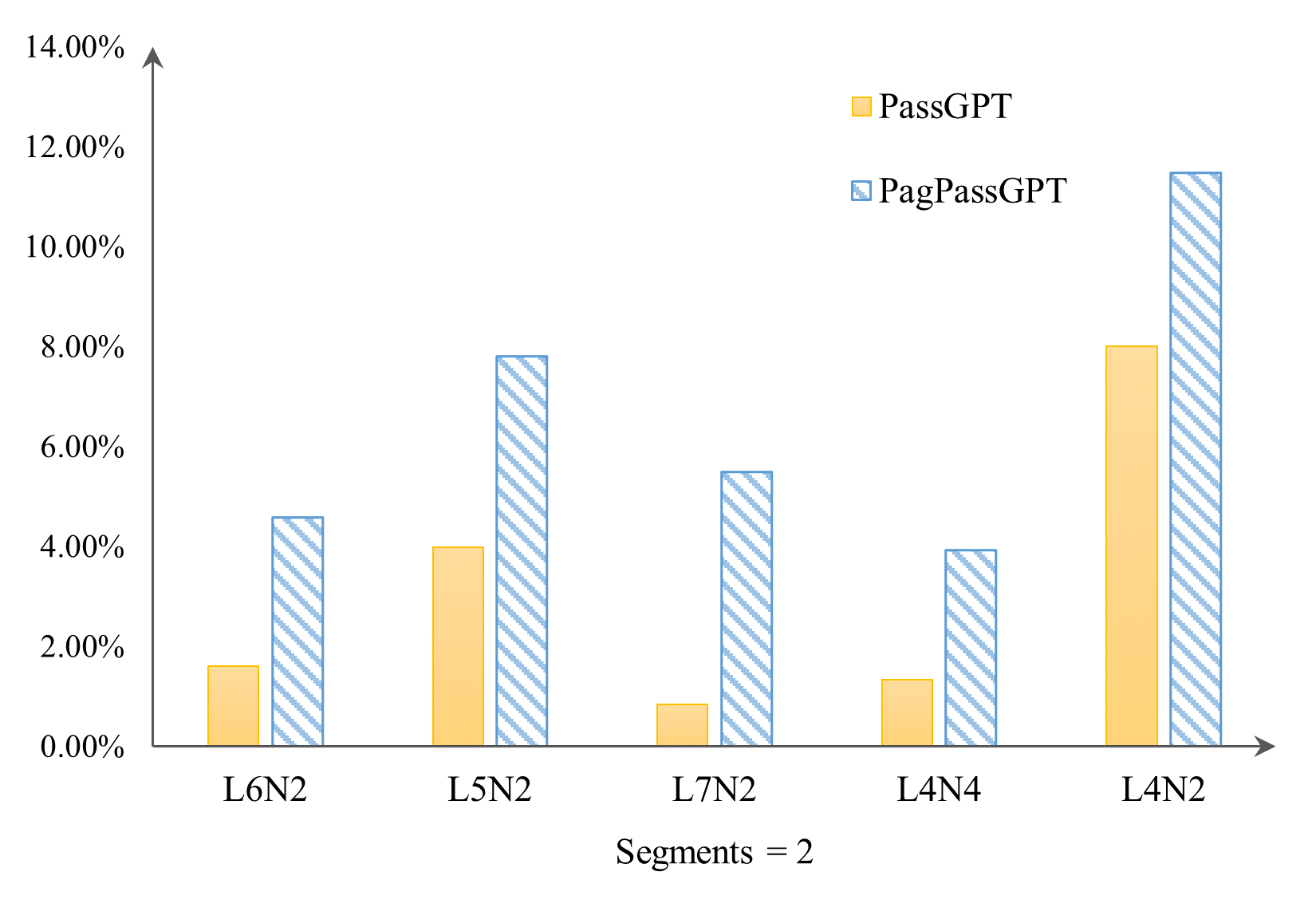}
        \label{fig:chunksize_2}}
        \hfil
        \subfloat[]{\includegraphics[width=0.9\columnwidth]{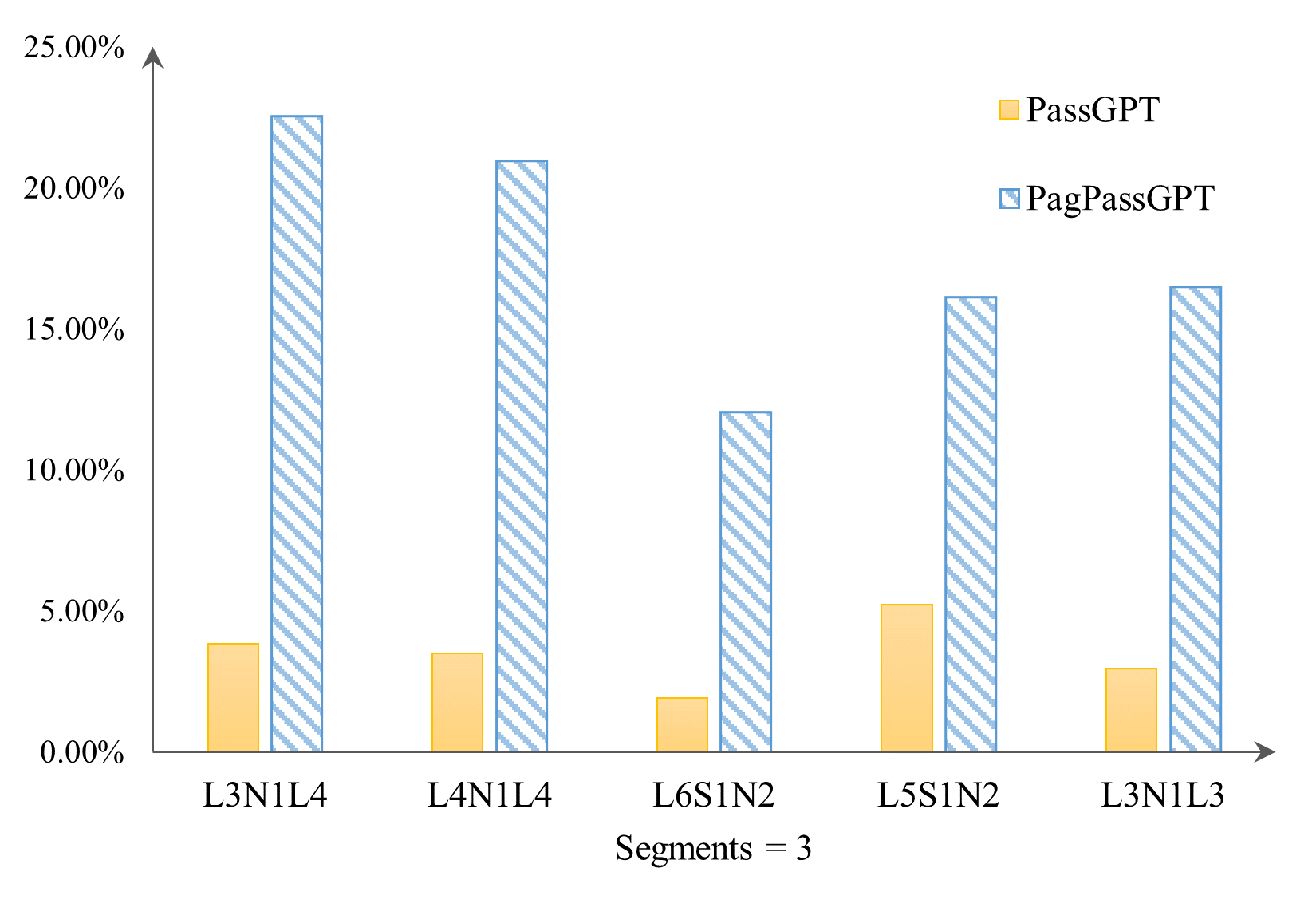}
        \label{fig:chunksize_3}}
        \hfil
        \subfloat[]{\includegraphics[width=0.9\columnwidth]{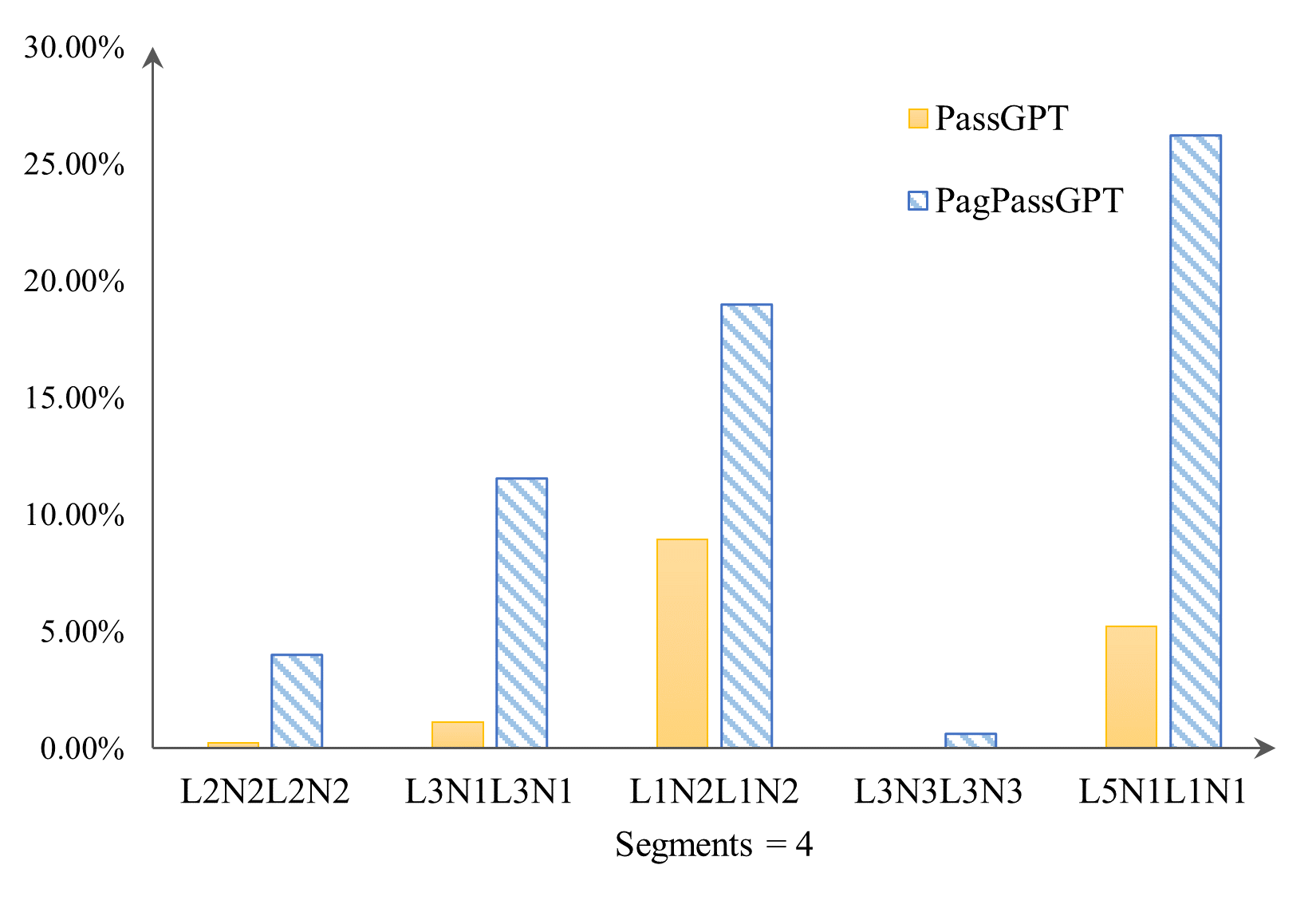}
        \label{fig:chunksize_4}}
        \hfil
        \subfloat[]{\includegraphics[width=0.9\columnwidth]{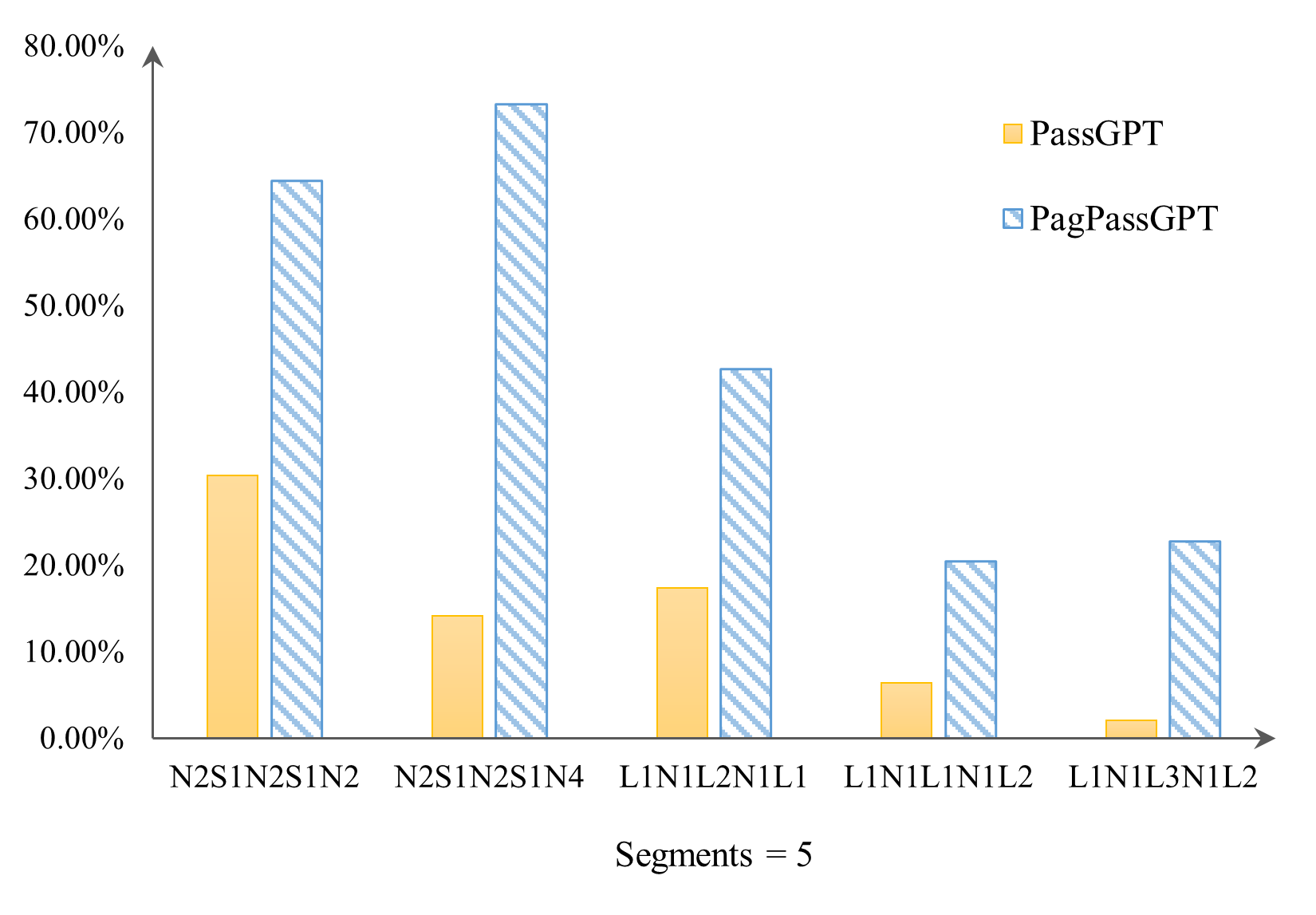}
        \label{fig:chunksize_5}}
        \hfil
        \subfloat[]{\includegraphics[width=0.9\columnwidth]{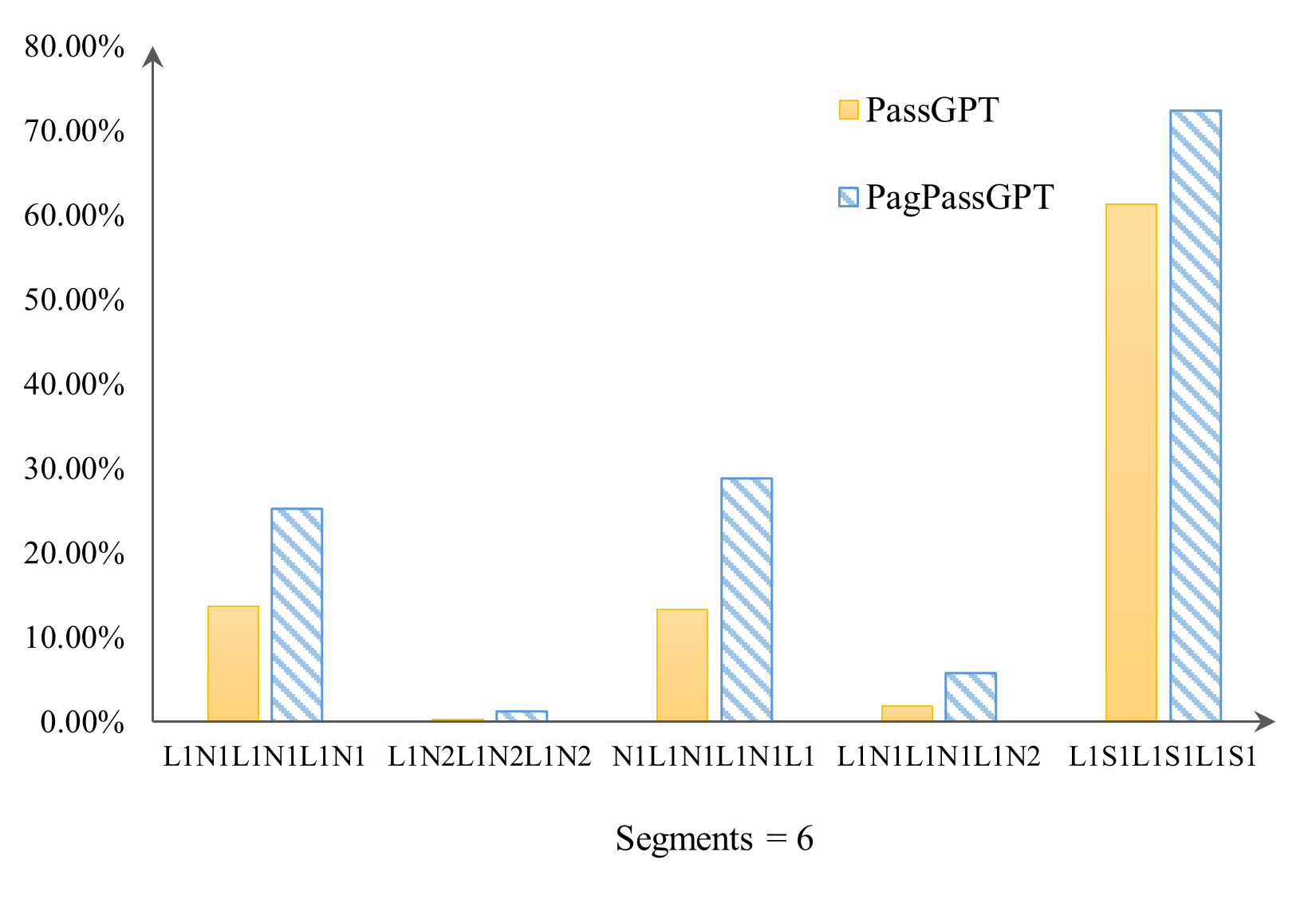}
        \label{fig:chunksize_6}}
    \caption{Compare $HR_P$ of PassGPT and PagPassGPT, $s\in [1,6]$ and $P$ from Top 5. The vertical axis represents the hit rate $HR_P$, while the horizontal axis represents different patterns.}
    \label{fig:chunksize=1-6}
\end{figure*}

As shown in Fig. \ref{fig:chunksize=1-12}, with an increasing number of pattern segments, PagPassGPT consistently outperforms PassGPT. 
For instance, when the number of segments is $1$, the distinction between PagPassGPT and PassGPT is less pronounced. 
As the number of segments increases to $5$, the gap reaches its peak, with the hit rates ($HR_s$) of 13.00\% for PassGPT and  40.54\% for PagPassGPT.
When the number of segments exceeds  $9$, the hit rate of PassGPT approaches zero. However, PagPassGPT continues to demonstrate its utility.
We further illustrate the details of the hit rate of each pattern in Fig. \ref{fig:chunksize=1-6}. 
For the convenience of presentation, we only show the top 5 patterns of each category from the first segment to the sixth segment.
PagPassGPT demonstrates a higher hit rate ($HR_P$) for almost all patterns compared to PassGPT. Particularly, PagPassGPT is capable of guessing passwords with challenging patterns, while PassGPT fails to make any correct guesses.

To illustrate the disparity in pattern guided guessing between PassGPT and PagPassGPT, we randomly select ten passwords generated by each model, adhering to the ``L5N2" and ``L5S1N2" patterns, as outlined in Table \ref{tab:passwords}.
From the table, it is evident that passwords generated by PassGPT tend to exhibit word truncation, particularly when English words are involved.
For example, in the password ``polic\#10", the word ``police" lacks the letter ``e" because PassGPT must insert a special character in its subsequent token to adhere to the pattern requirement when the generation process reaches ``e".
In contrast, PagPassGPT rarely encounters such issues since it considers not only the pattern requirement but also the model prediction.

\begin{table}[tbp]
    \centering
    \caption{Passwords generated in pattern guided guessing test by PassGPT and PagPassGPT}
    \begin{tabular}{cccc}
    \toprule
        \multicolumn{2}{c}{\textbf{PassGPT}} & \multicolumn{2}{c}{\textbf{PagPassGPT}} \\ 
        \cmidrule(lr){1-2}\cmidrule(lr){3-4}
        L5N2 & L5S1N2 & L5N2 & L5S1N2 \\ \midrule
        stlad10 & polic\#10 & Sissi11 & sweet@74 \\
        matth10 & kimmy@90 & Panda51 & shock-22 \\
        taken11 & SexyB@20 & manan83 & deivi\_23 \\
        Calis31 & summe\_23 & tammy04 & loveu.18 \\
        sexyb32 & lovef\$45 & venus19 & cheer\_11 \\
        myboo54 & missl!12 & Homie04 & devan+12 \\
        veraj19 & boxer'20 & DANNY32 & faces\$25 \\
        djuju69 & trees-27 & green02 & sweet!21 \\
        plesn11 & mayho\{19 & Lucky15 & shock-22 \\
        poonk92 & gordi\_21 & brick22 & ilove\$32 \\
    \bottomrule
    \end{tabular}
    \label{tab:passwords}
\end{table}

\subsection{Trawling Attack Test}
\label{sec:Trawling Attack Guessing}

To assess the performance of PagPassGPT in trawling attacks, we compare them with recent and relevant works, including PassGAN, VAEPass, PassFlow, and PassGPT. 
In particular, PagPassGPT employed two approaches for generation. The first one is that the input is only a single token \BOS.
All subsequent content, containing the pattern and the password, is autonomously generated by the model itself.
Another approach is assisted by D\&C-GEN and the threshold $T$ of D\&C-GEN is set to 4,000 determined based on the parallelism capability of the applied GPU.
For convenience in subsequent discussions, we use PagPassGPT-D\&C to denote PagPassGPT equipped with D\&C-GEN.

\subsubsection{Hit Rate}
The hit rate is the ratio of passwords generated by the model that match with passwords in the test set to the total number of passwords in the test set. 
Both the generated passwords and the passwords in the test set undergo a deduplication process, ensuring that duplicates are eliminated before evaluating the hit rate.
 This metric is a key indicator for evaluating the performance of a password guessing model.

As depicted in Table \ref{tab:TA hit rate}, PagPassGPT demonstrates a superior hit rate compared to other deep learning-based password guessing models. 
Furthermore, D\&C-GEN enhances this advantage, achieving a hit rate of 53.63\% at $10^9$ guesses, approximately 12\% higher than PassGPT.

\begin{table}[tbp]
    \centering
    \caption{Hit rates of different models in trawling attack test.}     
    \label{tab:TA hit rate}
    \begin{tabular}{lrrrr}
        \toprule
        Guess Num & $10^6$ & $10^7$ & $10^8$ & $10^9$ \\
        \cmidrule(lr){1-1}\cmidrule(lr){2-5}
        PassGAN & 0.80\% & 3.11\% & 8.24\% & 16.32\% \\
        VAEPass & 0.49\% & 2.24\% & 6.24\% & 12.23\% \\
        PassFlow & 0.26\% & 1.62\% & 7.03\% & 14.10\% \\
        PassGPT & 0.73\% & 5.60\% & 21.43\% & 41.93\% \\
        PagPassGPT & 1.00\% & 7.68\% & 27.23\% & 48.75\% \\
        PagPassGPT-D\&C & \textbf{1.05\%} & \textbf{8.48\%} & \textbf{31.38\%} & \textbf{53.63\%} \\ 
        \bottomrule
    \end{tabular}
\end{table}

\subsubsection{Repeat Rate}

\begin{figure}[tbp]
    \centering
    \includegraphics[width=1.0\columnwidth]{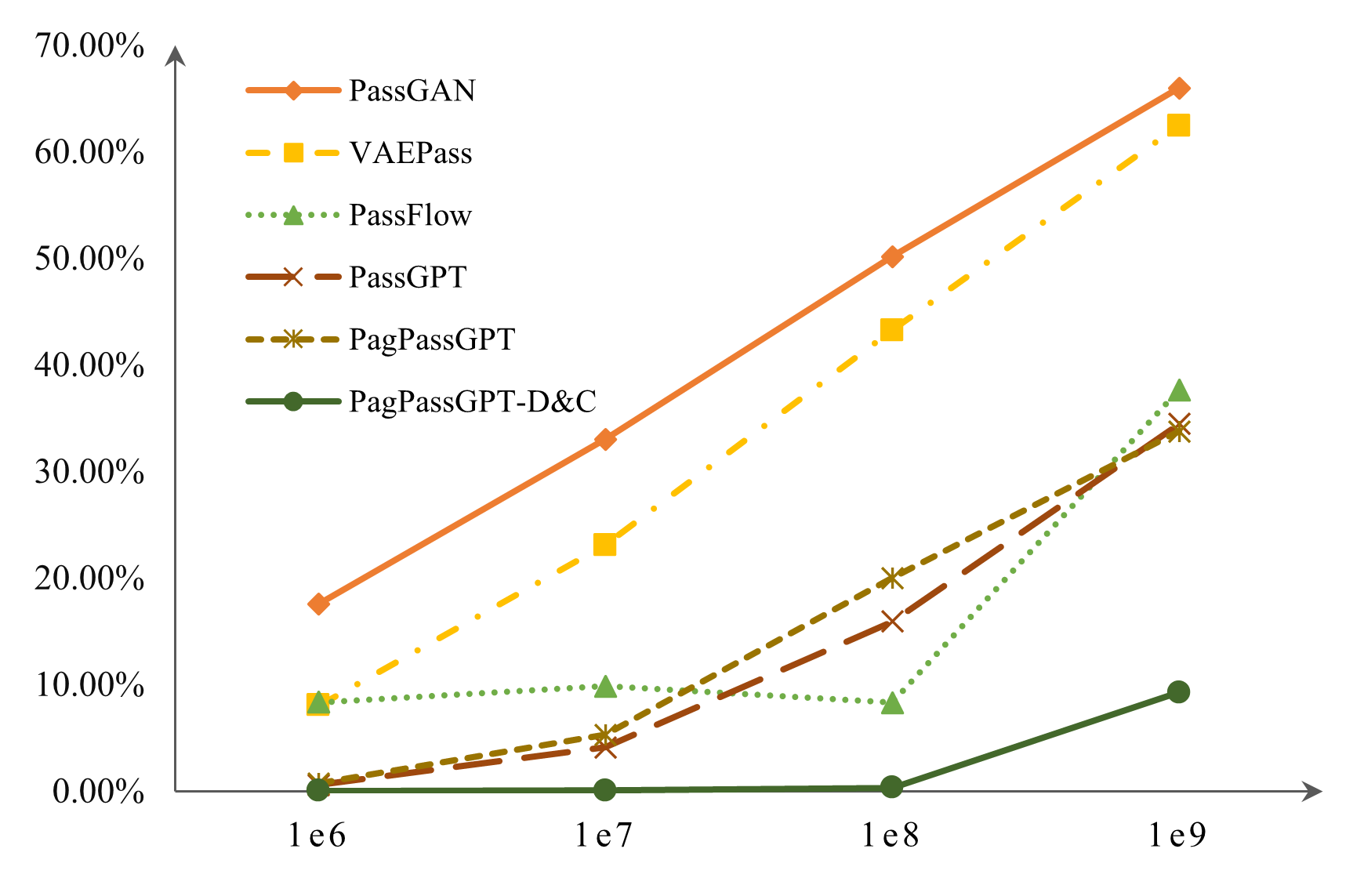}
    \caption{Repeat rates of passwords generated by different models.}
    \label{fig:repeat rate}
\end{figure}

The repeat rate reflects the percentage of duplicate passwords among those generated by a model. 
For the generated passwords and the passwords in the test set are all deduplicated, generating passwords that have been generated will not increase the hit rate.
Therefore, when the generation number has been set, a high repeat rate diminishes the effective diversity of generated passwords, potentially impacting the model's hit rate. Thus, monitoring the repeat rate is crucial in the context of trawling attacks.

As shown in Fig. \ref{fig:repeat rate}, as the number of generated passwords increases, our model shows a slower increase in repeat rate compared to other models. 
With the assistance of D\&C-GEN, PagPassGPT-D\&C achieves a repeat rate of 9.28\% with $10^9$ generated passwords. In contrast, PassGPT exhibits a repeat rate of 34.5\%, approximately 25\% higher than PagPassGPT-D\&C. All the remaining models have higher repeat rates than PassGPT.

\subsubsection{Length Distribution and Pattern Distribution}
To provide a more comprehensive understanding of PagPassGPT's effectiveness in terms of password quality, we conduct a detailed analysis of the length distribution and pattern distribution of the generated passwords. A closer alignment with the characteristics of the test set indicates superior performance.

In particular, following the configuration of PassGPT~\cite{passgpt},  we compare the length distribution and pattern distribution of $10^8$ passwords generated by various models, including PassGAN, VAEPass, PassFlow, and PagPassGPT.
PagPassGPT-D\&C, as it requires patterns as input and generates passwords guided by patterns, is excluded from the comparison.

The variance of distributions is assessed using both length distance and pattern distance. 
Both metrics are computed through the Euclidean distance between the distributions of generated password lengths and patterns and those found in the test set.

\begin{align}
\label{eq:length distribution}
    D_{length} = \left( \sum_{i=4}^{12} (Pr_{test}(L_i)-Pr_{model}(L_i))^2 \right)^{1/2}  \\ 
\label{eq:pattern distribution}
    D_{pattern} = \left(\sum_{i=1}^{150} (Pr_{test}(P_i)-Pr_{model}(P_i))^2 \right)^{1/2} 
\end{align}

For length distance, as defined in \eqref{eq:length distribution}, we consider passwords with 4 to 12 characters, which aligns with the data cleaning process. $Pr_{test}$ represents the probability distribution from the test set, and $Pr_{model}$ represents the distribution from the model output.
For pattern distance, as illustrated in \eqref{eq:pattern distribution}, we focus on the distribution of the top 150 common patterns in the test set, as their cumulative probability exceeds 90\% and effectively represents the overall pattern distribution of generated passwords. Similarly, we conduct the same analysis on the passwords generated by the models.

\begin{table}[tbp]
    \centering
    \caption{Length distances and pattern distances between passwords generated by different models and the test set.}
    \begin{tabular}{lrr}
    \toprule
       Model & Length Distance & Pattern Distance \\
       \cmidrule(lr){1-1}\cmidrule(lr){2-3}
       PassGAN    & 9.20\%     & 6.00\% \\ 
       VAEPass    & 5.84\%     & 5.75\% \\
       PassFlow   & 50.61\%    & 13.62\% \\
       PassGPT    & 8.49\%     & 4.16\% \\
       PagPassGPT & \textbf{4.78\%} & \textbf{2.79\%} \\ 
       \bottomrule
    \end{tabular}
    \label{tab:distributions}
\end{table}

\begin{figure}[tbp]
    \centering
    \includegraphics[width=1.0\columnwidth]{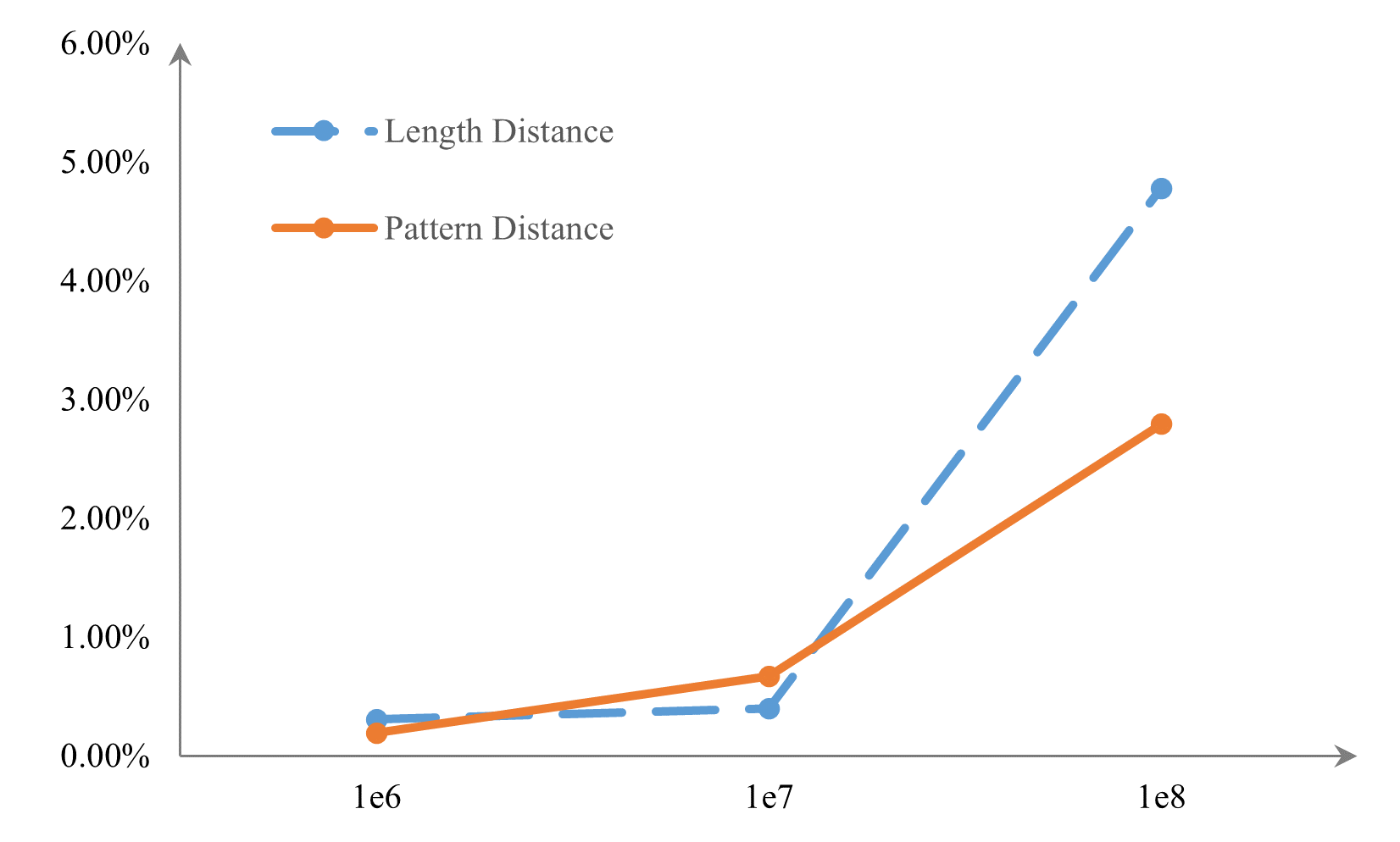}
    \caption{Length distances and pattern distances of PagPassGPT.}
    \label{fig:length distance changing}
\end{figure}

As illustrated in Table \ref{tab:distributions}, PagPassGPT exhibits its distribution closest to the test set when compared with other models. 
The length distance of PagPassGPT is 4.78\%, roughly half of the length distance of PassGPT. 
Similarly, the pattern distance of PagPassGPT is  2.79\%, compared to PassGPT's pattern distance of 4.16\%.
To better understand PagPassGPT,  we conducted a further analysis of the length distances and pattern distances on different numbers of passwords generated by PagPassGPT.
As illustrated in Fig. \ref{fig:length distance changing}, both distances increase with the growing number of passwords. 
Especially, the distances increase significantly from 1e7 to 1e8
due to the rise of the repeat rate.

\begin{table}[tbp]
    \centering
    \caption{Hit rates of different models in cross-site attack test.
    }
    \label{tab:cross-site}    
    \begin{tabular}{lrrr}
    \toprule
        \multicolumn{4}{c}{Trained on Rockyou}   \\ 
        \midrule
        Model & phpBB & MySpace & Yahoo! \\
        \cmidrule(lr){1-1}\cmidrule(lr){2-4}
        PassGPT  & 31.30\% & 43.13\% & 28.79\%  \\
        PagPassGPT & 40.13\% & 53.79\% & 36.72\%   \\ 
        PagPassGPT-D\&C  & \textbf{43.48\%} & \textbf{56.76\%} & \textbf{39.47\%}   \\ 
        \midrule
        \multicolumn{4}{c}{Trained on LinkedIn} \\ 
        \midrule
        Model & phpBB & MySpace & Yahoo! \\
        \cmidrule(lr){1-1}\cmidrule(lr){2-4}
        PassGPT  &  28.45\% & 35.69\% & 28.94\%  \\
        PagPassGPT &  35.38\% & 45.20\% & 35.81\%  \\ 
        PagPassGPT-D\&C  & \textbf{44.16\%} & \textbf{50.30\%} & \textbf{39.13\%}  \\ 
        \bottomrule
    \end{tabular}
\end{table}

\subsection{Cross-Site Attack Test}
\label{sec:cross-site attack}

To evaluate the generality of the proposed model, we conduct a cross-site attack test. 
We first train the most recent PassGPT and the proposed PagPassGPT
on Rockyou and LinkedIn independently. 
Then we evaluate them by testing the hit rates of $10^8$ passwords on other datasets and the result of hit rates is shown in Table \ref{tab:cross-site}. 
PassGAN, VAEPass, and PassFlow are excluded from the comparison since their hit rates show a significant gap with PassGPT and PagPassGPT at $10^8$ guesses in the trawling test. Specifically, their hit rates are less than 10\% while the hit rates of both PassGPT and PagPassGPT are over 20\% as shown in Table \ref{tab:TA hit rate}.

From Table \ref{tab:cross-site}, it is evident that PagPassGPT demonstrates better generalization compared to PassGPT. Moreover, PagPassGPT-D\&C is able to further enhance the performance by 3\% to 10\%. 
Compared to PassGPT,  PagPassGPT-D\&C achieves an 11\% to 16\% higher hit rate.

\section{Limitations and Discussion}
\label{sec:discussion}
In this section, we will discuss the limitations of the proposed models and insights about the password generation algorithm. 

\textbf{Limitations}. 
The present version of PagPassGPT exhibits limitations in diversity for pattern guided guessing,  solely supporting patterns extracted by PCFG.
In addition,  in the context of trawling attacks, PagPassGPT generates passwords within a restricted length range, capped at 12 characters. 
Owing to the use of position encoding in GPT, the input window size and the acceptable output text length are predetermined once the training parameters are set. Nevertheless, training a new model for generating longer passwords is a straightforward process, accomplished by extending the input window. Similarly, if we need to extend the search space, a new model for accepting more characters should be trained just by adding new characters into the vocabulary of the tokenizer.
Finally, while D\&C-GEN has improved performance and reduced the repeat rate, it also extends the required time for division.
A small threshold for dividing a task leads to more divisions of guessing tasks and a lower repeat rate. Taking into account memory consumption and the maximum available threads, we can establish the maximum number of parallel subtasks and determine the optimal threshold accordingly.

\textbf{Insights}. 
We believe that an effective password guessing model can be considered as two parts:  password knowledge extraction and password generation using obtained knowledge.
These two components are complementary to each other. 
In prior research, attention was primarily directed towards the first part of the password modeling,
often overlooking the significance of the second part. 
Without a well-designed second part, the extracted knowledge cannot be fully utilized to generate passwords.

\section{Conclusions}
\label{sec:conclusions}
In this paper, we introduced PagPassGPT, a password guessing model, and D\&C-GEN, a password generation algorithm. PagPassGPT excels in producing high-quality passwords with pattern requirements, and when coupled with D\&C-GEN, our model demonstrates outstanding performance in both pattern guided guessing, trawling attack guessing, and cross-site attack guessing, showcasing higher hit rates and lower repeat rates. Furthermore, we discussed the limitations of our solutions and underscored the significance of the generation algorithm in password guessing, identifying it as a possible focal point for future research.

\normalem
\bibliographystyle{IEEEtran}
\bibliography{main}

\end{document}